# Determination of the Sign of *g* factors for Conduction Electrons Using Time-resolved Kerr Rotation

C. L. Yang [1,2], Junfeng Dai [1], W. K. Ge [2], Xiaodong Cui [1,*]

1. *Department of Physics, The University of Hong Kong, Hong Kong, China*
2. *School of Physics and Engineering, Sun Yat-sen University, 510275 Guangzhou, China*

**Abstract:** The knowledge of electron g factor is essential for spin manipulation in the field of spintronics and quantum computing. While there exist technical difficulties in determining the sign of g factor in semiconductors by the established magneto-optical spectroscopic methods. We develop a time resolved Kerr rotation technique to precisely measure the sign and the amplitude of electron g factor in semiconductors.

* Email: xdcui@hku.hk

The emerging technology of spintronics and quantum computing utilizes the electron spin degree of freedom. One approach toward the spin manipulation is to tune electron's *g* factor (Landé g factor) which associates magnetic dipole momenta with electron spin momenta (or total angular momenta), by engineering semiconductor structures. The knowledge of electron *g* factor is essential for many aspects. Charactering electron *g* factor in semiconductor structures has been of great importance in the research of spintronics and quantum computation. Traditionally, electron *g* factor in semiconductors has been measured by electron spin resonance spectroscopy (ESR) and magneto-transport techniques which probe properties of electrons at Fermi level. However, the application of ESR on semiconductor heterostructures has been rarely reported due to experimental difficulties.[1,2] The magneto-transport measurements request relatively high criteria on sample quality. Besides, it is often ambiguous for both techniques to determine the sign of *g* factor. In addition to the ESR and magneto-transport techniques, magneto-optical spectroscopic techniques particularly magneto-photoluminescence and Kerr/Farrady rotation spectroscopy provide plenty of information on the electron spin in semiconductor structures. Magneto-photoluminescence probes the electron spin by analyzing the circular polarization of the photoluminescence (PL) under magnetic field.[3] Since the electron (hole) *g* factor of semiconductors is deduced from the exciton's or electron-hole pair's *g* factor, which is the sum of an electron and a hole *g* factor, the correct information of the electron (hole) g factor will be only possible if the counter part is known. Very recently Kalevich *et al* reported a method of measuring the sign of the electron *g* factor in semiconductor quantum wells (QWs) from magneto-PL spectra.[4] The method utilizes asymmetry in the depolarization of PL spectra in a transverse magnetic field under the Hanle effect. Whereas, as the PL in semiconductors usually comes from complex origins, the magneto-spectra often give ambiguous information on the conduction electron spin if shallow defect states are involved, which frequently occurs in low temperature PL.

Optical Kerr rotation and Faraday rotation spectroscopy are found to be simple yet powerful to accurately probe the spin polarization in a wide range of semiconductor structures. Given the fact that Kerr rotation signal is dominated by the spin information of electrons in conduction band and the holes in valence band, the *g* factor determined in this way will be more reliable if the spin information of the conduction electrons is concerned. To the authors' knowledge, Kerr rotation spectroscopy has been only used to determine the magnitude of electron g factors.[5] In this letter we report two methods utilizing the time-resolved Kerr rotation spectroscopy to determine both the magnitude and the sign of the *g* factor in various semiconductor structures: GaAs thin film, GaAs two dimensional electron gas (2DEG) and GaNAs/GaAs quantum wells (QW)

The phenomenological mechanism of our experiments is sketched in Figure 1(a): Electron spin polarization is injected by the oblique incidence of the circularly polarized pump light, and its time evolution of the spin polarization along z direction under transverse external magnetic field is monitored by the back reflected linearly polarized probe light at normal direction (*z*-axis). In this scenario, spin polarized electrons in the magnetic field experience a magnetic torque $\vec{T} = \vec{\mu} \times \vec{B} = (-g\mu_B/\hbar)\vec{S} \times \vec{B}$ where $\vec{\mu}$ is the magnetic moment, $\vec{B}$ is the transverse magnetic field, $g$ is the electron Landé $g$ factor, $\mu_B$ is the Bohr magneton, $\hbar$ is the Plank constant and $\vec{S}$ is the spin angular momentum. Considering of the spin relaxation and dephasing, the average spin of photo-generated electrons under the magnetic torque could be described by the equation $\frac{d\vec{S}}{dt} = \vec{\omega} \times \vec{S} - \frac{\vec{S}}{\tau_S}$ where $\vec{\omega} = g\mu_B \vec{B}/\hbar$ is the Larmor precession frequency and $\tau_S$ is the spin lifetime. It shows that the direction of the spin precession will be determined by both the sign of the *g* factor and the orientation of the magnetic field. As shown in Fig. 1(b), spin processes either clockwise (for g<0) or counter-clockwise (for g>0) in the plane perpendicular to the magnetic field driven by the magnetic torque. Consequently the sign of g factor could be well resolved from the time evolution of spin polarization along z direction ($S_z$) which the time resolved Kerr rotation can clearly tell.

In the experimental setup as sketched in Fig. 1(a), spin oscillation measured at the probe direction (z axis) under a transverse external magnetic field can be well described by $S_z(t) = S_0 \exp(-t/\tau_S)\cos(g\mu_B Bt/\hbar + \varphi)$, where $S_0$ is the initial amplitude of the spin and $\varphi$ is the phase shift related to the angle $\alpha$ between the pump and the probe beams. In our particular setup, $\varphi$ is about $\pm\pi/27$ by taking $\alpha = 25^0$ and refractive index n = 3.6 for GaAs, depending on the spin precession direction: $\varphi < 0$ for clockwise spin precession or $\varphi > 0$ for counter clockwise. Both pump and probe beams are from a tunable mode-locked Ti:Sapphire laser with a pulse width about 150 femtoseconds and a repetition rate of 80 MHz. The pump beam size is about 30 μm in focus and the size of the probe beam is tuned to be a little bit smaller than the former. The typical excitation powers are of 2 mW for the pump and 0.4 mW for the probe beams, respectively. To get excellent signal-to-noise ratio, a double lock-in detection technique was employed with the amplitude modulation of the probe beam at 115 Hz with an optical chopper and the polarization modulation of the pump beam at 50KHz by a photoelastic modulator. The measurements were carried out in a magneto-optical cryostat. Samples include GaAs thin film, GaAs 2DEG and GaNAs/GaAs QWs.

Figure 2 shows the experimental time-resolved Kerr rotation data for (a) GaAs thin film, (b) GaAs 2DEG and (c) GaAsN/GaAs QW with nitrogen composition of 1.5%. (a) and (b) were measured at T = 5 K and (c) was at T = 220 K. Due to the fast spin relaxation of holes, the observed Kerr rotation signal is usually dominated by the spin relaxation of electrons. The damped oscillations of the Kerr signals in Figure 2 clearly show that the injected electron spin precesses around the magnetic field during its lifetime. The g factors extracted by the oscillation period are 0.42, 0.36 and 0.97 for (a), (b) and (c) respectively. To get the sign of the g factor, one has to examine the difference between the Kerr data with different magnetic field orientation. The curves indicate that there are time shifts for the Kerr signals in all the samples when the magnetic field is reversed. In GaAs thin film and 2DEG, the curve at $B>0$ lags behind that at $B<0$, suggesting $g<0$ as depicted in Fig 1(b), which is consistent with the well known negative g factor for GaAs. For the GaAsN/GaAs QW, the sign of the phase shift is reversed, suggesting the positive sign of the g factor, which is also in good agreement with previous [6].

To further demonstrate that the time shift of the spin oscillation from negative to positive magnetic field is only determined by the initial phase shift between clockwise and counter clockwise spin precession, we have measured the Kerr rotation on the GaAsN/GaAs QW sample at (a) B = ±0.3 T, (b) B = ±0.5 T and (c) B = ±0.9 T at temperature T = 5K, as shown in Figure 3. It clearly indicates that the time shift between positive and negative fields is getting smaller as the field increases. But the phase shifts between the positive and negative fields in these three cases are found to be a constant of about 0.22 and is independent on the magnetic field. This phase shift is in good with agreement with the value of $2\varphi = 2(\pi/27) = 0.232$ determined by the experimental geometry.

We found that the sign of g factor can also be determined by monitoring Kerr rotation signal in a sweeping magnetic field at a fixed delay time from the coherent spin injection. Figure 4 shows the magnetic field dependence of the Kerr rotation signal at a delay time of 170ps for (a) GaAs thin film, (b) GaAs 2DEGs and (c) GaAsN/GaAs QW. The Kerr rotation signals oscillate against magnetic field, while the amplitude does not decay due to the fixed time. Clearly a shift of the curves occurs when the direction of the magnetic field is reversed. In both Figure 4(a) and (b) the curves are rightward shifted if the magnetic fields switch from negative to positive, Whereas the opposite shift occurs at GaAsN/GaAs QW as in Figure 4(c). The remarkably different behavior of GaAsN/GaAs QW implies the different sign of g factor of GaAsN/GaAs QW from others. It is worth mentioning that this method of determining the sign of g factors is often expected to gives more robust results than the method used in Fig. 2, considering of a time shift of ~10 *ps* in Figure 2 versus a field shift of ~1000 Gauss in Figure 4.

In conclusion, we develop two experimental configurations of time-resolved Kerr rotation spectroscopy for electron g factor study. The magnitude and the sign of g factor of conduction electrons in semiconductors can be unambiguously determined both by the spin evolution under transverse magnetic field and by scanning magnetic field at a fixed time delay.

Acknowledgement
The work was supported by Hong Kong General Research Fund under HKU 701308P and Collaborative Research Fund under HKU10/CRF/08, China NSF under Grants No. 60706021, No. 10874248, and No. 60876066.

References

1. M. Dobers, K. von Klitzing and G. Weimann, Phys. Rev. B 38, 5453 (1988).
2. M. Krapf, G. Denninger, H. Pascher, G. Weimann, and W. Schlapp, Solid State Commun. 74, 1141 (1990).
3. M. J. Snelling, E. Blackwood, C. J. Mcdonagh, R. T. Harley and C. T. B. Foxon, Phys. Rev. B 45, 3922 (1992).
4. V. K. Kalevich, B. P. Zakharchenya, K. V. Kavokin, A. V. Petrov, P. Le Jeune, X. Marie, D. Robart, T. Amand, J. Barrau, and M. Brousseau, Phys. Solid State 39, 681 (1997).
5. J. A. Gupta, D. D. Awschalom, X. Peng and A. P. Alivisatos, Phys. Rev. B 59, R10 421 (1999).
6. V. K. Kalevich, E. L. Ivchenko, A. Yu Shiryaev, M. M. Afanasiev, A. Yu Egorov, M. Ikezawa and Y. Masumoto，Semicond. Sci. Technol. 23, 114008 (2008)

**Figure Captions**

**Figure 1(a)** Schematic diagram of the experimental configuration for Kerr rotation setup. The magnetic field is in the sample plane. The probe light is with a back-reflected geometry and the pump beam is with oblique incidence at an angle of $\alpha = 20^0$ away from the probe beam. **(b)** shows that spin will experience either clockwise (for g<0) or counter-clockwise (for g>0) rotation in the plane perpendicular to the magnetic field due to the magnetic torque. **(c)** indicates that the change of the magnetic field direction will also reverse the direction of the spin precession, which is helpful to determine the sign of the g factor.

**Figure 2** The time-resolved Kerr rotation data for **(a)** GaAs thin film, **(b)** GaAs two dimensional electron gas and **(c)** GaAsN/GaAs quantum well with nitrogen composition of 1.5%. (a) and (b) at T = 5 K and (c) at T = 220 K.

**Figure 3** The Kerr rotation data on the GaAsN/GaAs QW sample measured at **(a)** B = 0.3 T, **(b)** B = 0.5 T and **(c)** B = 0.9 T. The temperature is at T = 5 K.

**Figure 4** The magnetic field dependence of the Kerr rotation signal at fixed delay time of 170 ps for **(a)** GaAs thin film, **(b)** GaAs 2DEGs and **(c)** GaAsN/GaAs QW.

**Figure 1.**

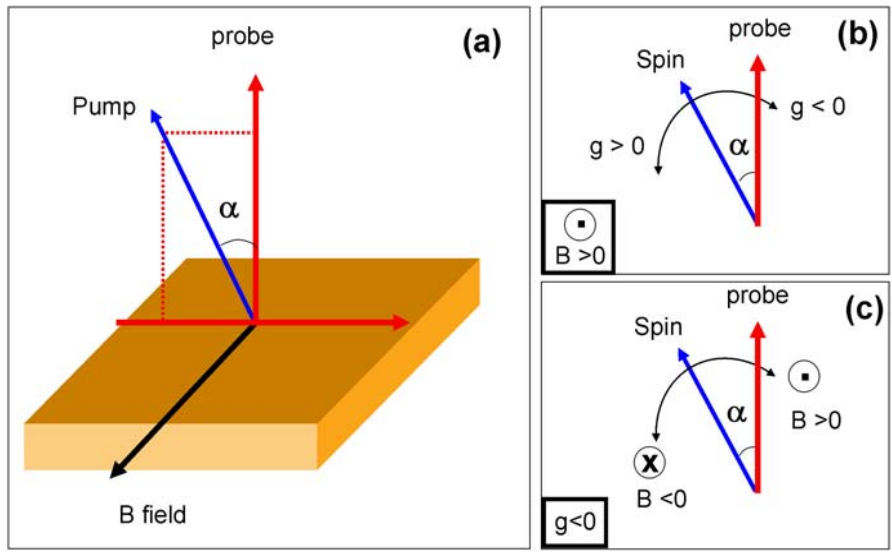

Figure 2.

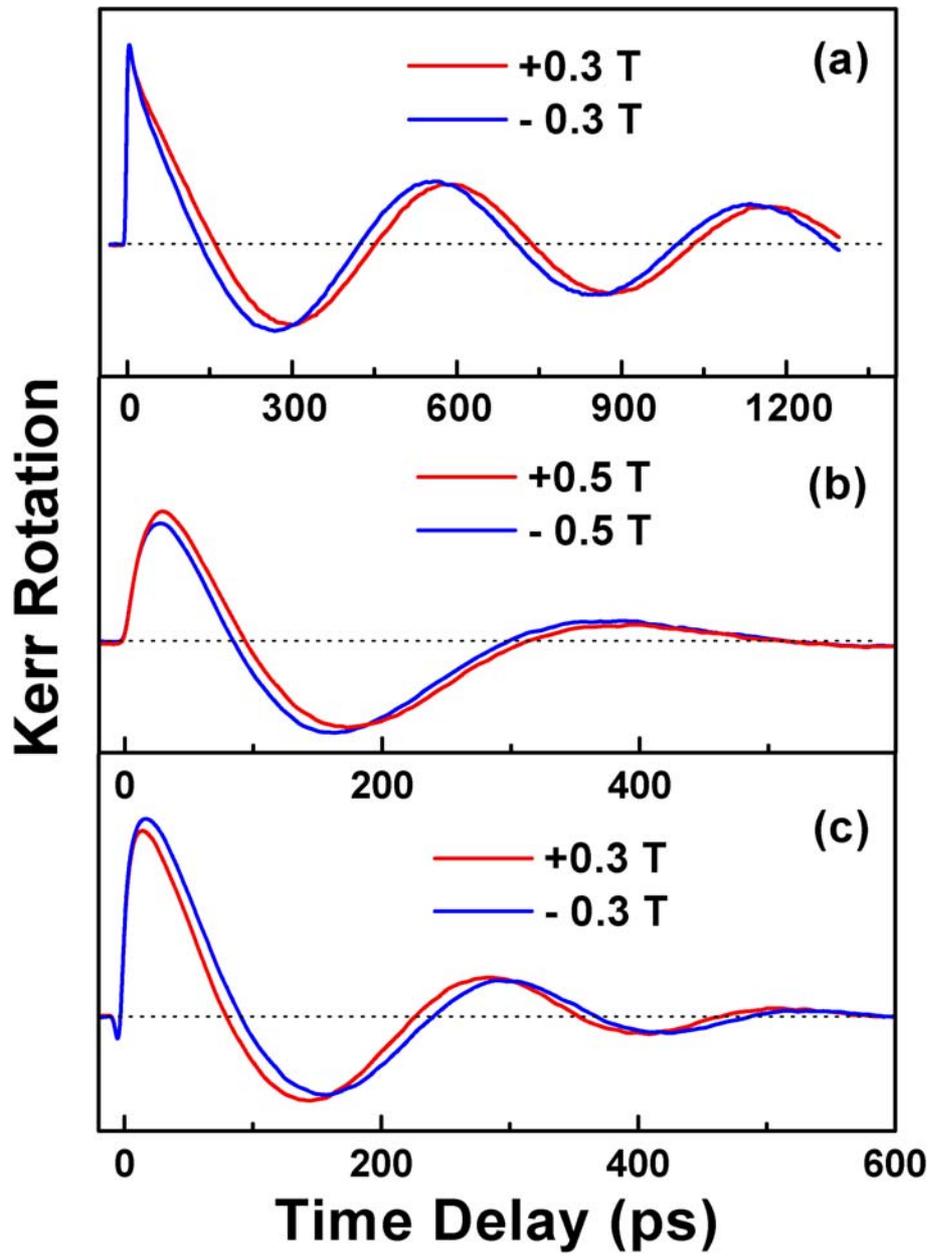

Figure 3.

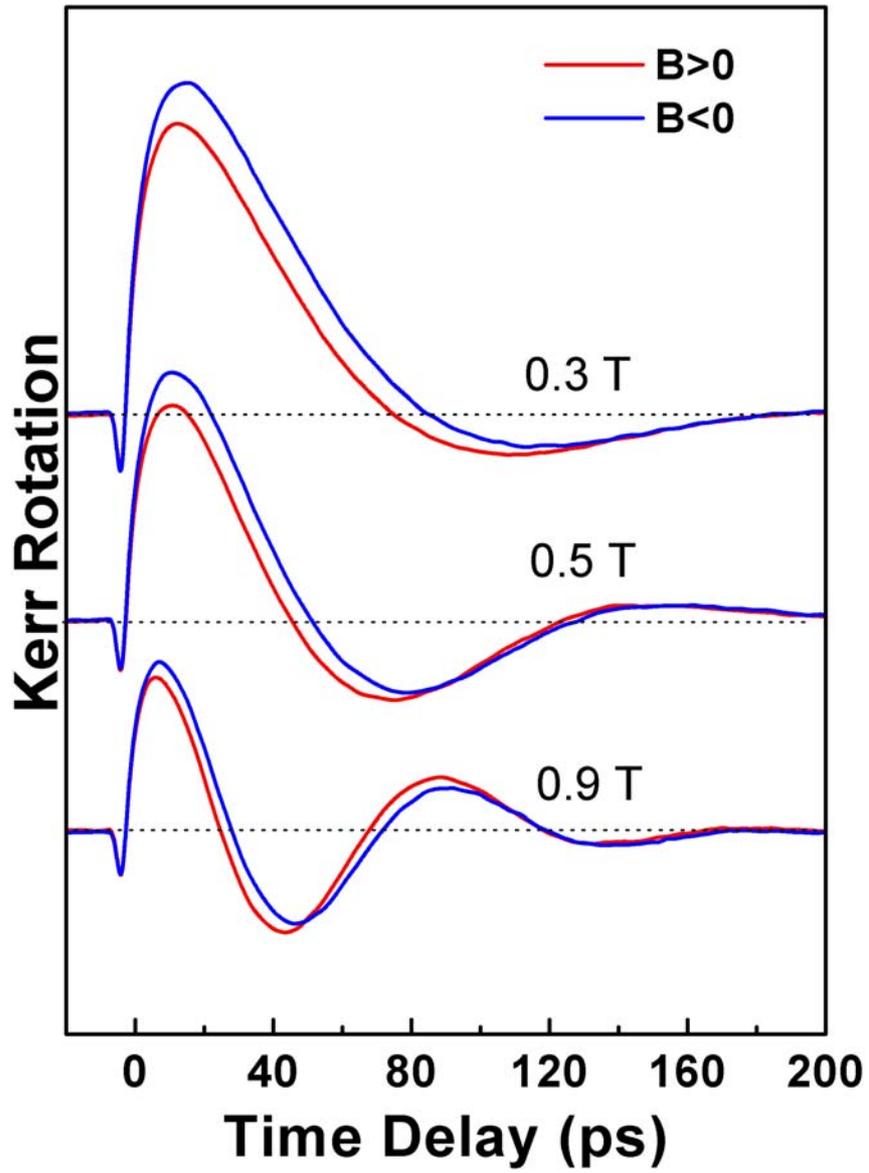

Figure 4.

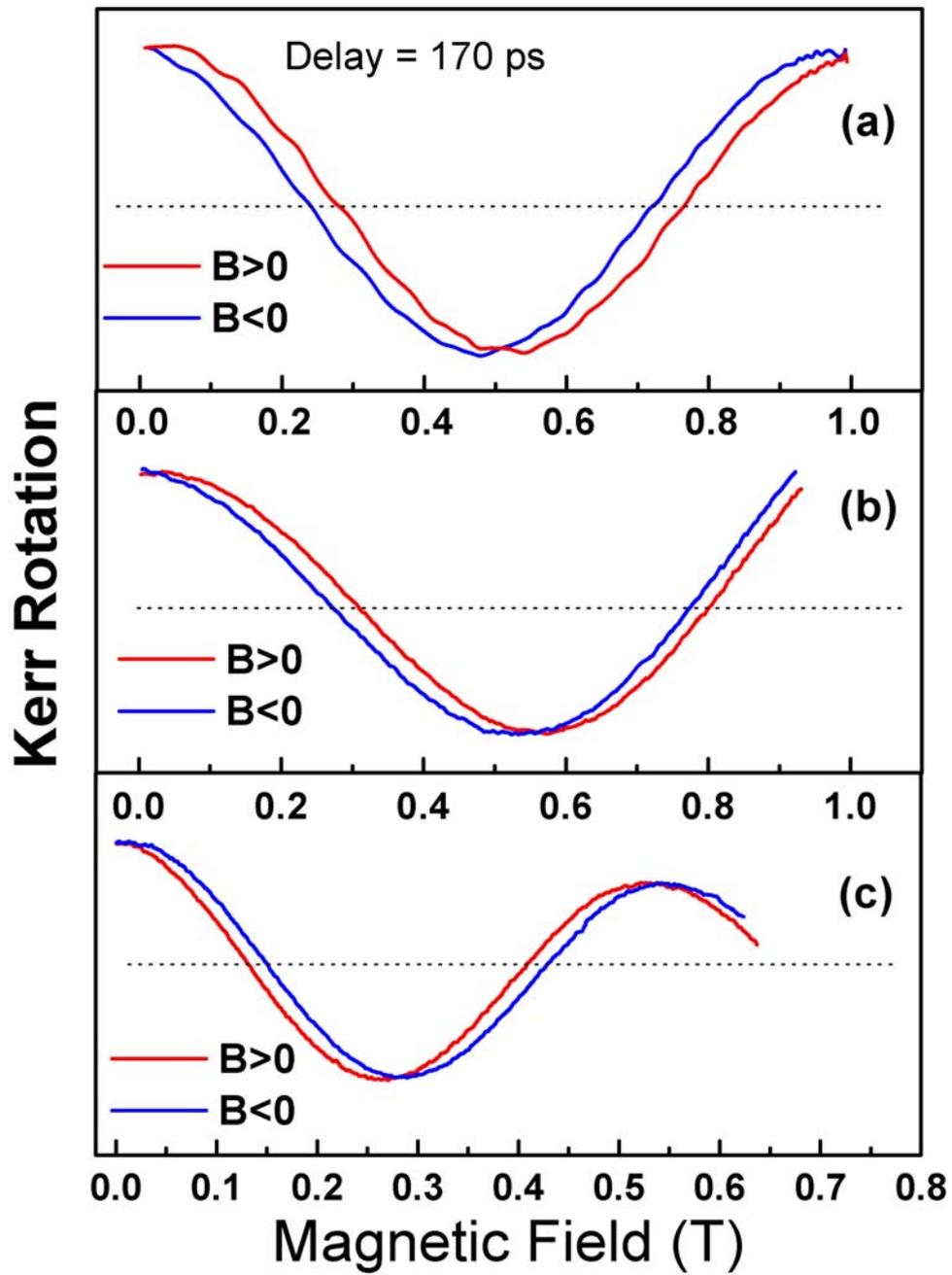